\documentclass[final,1p,times,twocolumn]{elsarticle}

\usepackage[usernames,dvipsnames]{xcolor}

\usepackage[ruled,vlined]{algorithm2e}
\usepackage{graphicx}
\usepackage{float}
\usepackage[english]{babel}
\usepackage{amssymb}
\usepackage[normalem]{ulem}
\usepackage{amsmath}
\usepackage{amsthm}
\usepackage{amsfonts}
\usepackage{enumerate}
\usepackage{multirow}
\usepackage{hyperref}
\usepackage{pdfpages}

\newcommand\nnfootnote[1]{%
  \begin{NoHyper}
  \renewcommand\thefootnote{}\footnote{#1}%
  \addtocounter{footnote}{-1}%
  \end{NoHyper}
}

\graphicspath{{images/}}

\begin{document}

\begin{frontmatter}


\title{A Fully Implicit Method for Robust \\Frictional Contact Handling in Elastic Rods}



\author[a]{Dezhong Tong$^{*,}$}
\author[b]{Andrew Choi$^{*,}$}
\author[c]{Jungseock Joo}
\author[a]{M. Khalid Jawed$^{\dagger,}$}

\nnfootnote{$^*$ Equal contribution.}
\nnfootnote{$^\dagger$ Corresponding author. email: \url{khalidjm@seas.ucla.edu}}

\address[a]{Department of Mechanical and Aerospace Engineering,}
\address[b]{Department of Computer Science,}
\address[c]{Department of Communication, \\ University of California, Los Angeles, \\ Los Angeles, California 90095, United States}


\begin{abstract}
{\it 
Accurate frictional contact is critical in simulating the assembly of rod-like structures in the practical world, such as knots, hairs, flagella, and more. Due to their high geometric nonlinearity and elasticity, rod-on-rod contact remains a challenging problem tackled by researchers in both computational mechanics and computer graphics.
Typically, frictional contact is regarded as constraints for the equations of motions of a system. Such constraints are often computed independently at every time step in a dynamic simulation, thus slowing down the simulation and possibly introducing numerical convergence issues.
This paper proposes a fully implicit penalty-based frictional contact method, Implicit Contact Model (IMC), that efficiently and robustly captures accurate frictional contact responses. 
We showcase our algorithm's performance in achieving visually realistic results for the challenging and novel contact scenario of flagella bundling in fluid medium, a significant phenomenon in biology that motivates novel engineering applications in soft robotics.
In addition to this, we offer a side-by-side comparison with Incremental Potential Contact (IPC), a state-of-the-art contact handling algorithm.
We show that IMC possesses comparable performance to IPC while converging at a faster rate.


}
\end{abstract}


\end{frontmatter}


\section{Introduction}
\label{sec:intro}

Throughout human history, flexible filamentary structures have been essential to human society, serving various purposes such as fastening, sailing, climbing, weaving, and hunting.
As our understanding of material properties improved, so did our ability to engineer rods with enhanced material properties (e.g., flexibility, strength, and resilience).
This in turn has resulted in the need to study and better understand the complicated mechanics of filaments.
Thus, several previous works have sought out to understand the various mechanics of rod-like structures including the deployment of rods~\cite{jawed2014coiling,jawed2014pattern,jawed2015geometric}, elastic gridshells~\cite{baek2018form,panetta2019x,baek2019rigidity}, plant growth~\cite{goriely2006mechanics}, knots~\cite{jawed2015untangling,audoly2007elastic, patil2020topological,patil2020discharging, grandgeorge2021mechanics}, and propulsion of bacterial flagella~\cite{jawed2015propulsion,jawed2016deformation,jawed2017dynamics}.
\begin{figure}[!t]
    \centering
    \includegraphics[width=\textwidth]{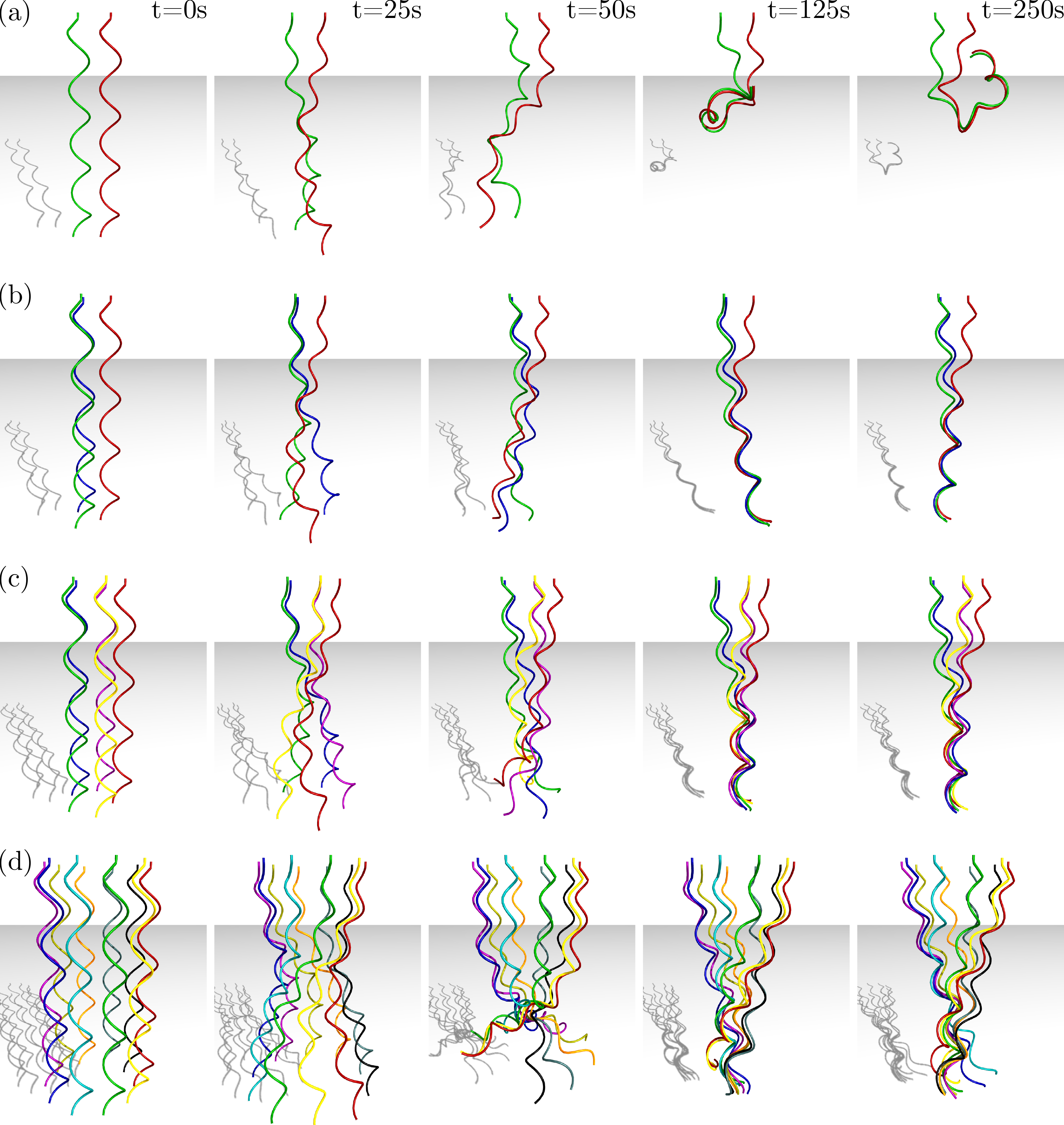}
    \caption{Rendered snapshots of flagella bundling with varying amounts of flagella. Rows contain (a) $M=2$, (b) $M=3$, (c) $M=5$, and (d) $M=10$ flagella. Each column indicates the flagella configuration at the moment of time indicated in the top row.}
    \label{fig:flagella_frames}
\end{figure}

With real-world experiments being costly and tedious to implement, the need for accurate physics-based numerical simulations is prevalent.
Such simulations not only allow for advanced mechanics-based study, they also open up the avenue to challenging problems in robotics ranging from simulating soft robot dynamics to closing the sim2real gap for deformable material manipulation.
In recent times, discrete differential geometry-based (DDG-based) simulations have shown surprisingly successful performance in capturing the nonlinear mechanical behaviors of rod structures~\cite{bergou2008discrete,bergou2010discrete}.
However, frictional contact handling still lacks a descriptive understanding.



Indeed, frictional contact formulations are usually diverse and based on the scaling of the system and/or the physical scenario. In this manuscript, we mainly focus on Coulomb friction, an adequate approximation of dry friction. Note that Coulomb friction degrades when contacted surfaces are conjoined. For engineering problems in which cohesion is important (e.g., cohesive granular media simulation~\cite{mandal2020insights, bertrand2005based, thakur2014micromechanical}), a more elaborated contact theory such as Johnson-Kendall-Roberts (JKR), Derjaguin–Muller–Toporov (DMT), or Maugis models~\cite{johnson1971surface, johnson1997adhesion, maugis1992adhesion} are required. Some prior works use these elastic cohesive contact models to simulate incipient sliding of cohesive contacts~\cite{borri2001incipient, gao2006nanoscale}. Despite this, Coulomb friction is still the de facto friction model for non-cohesive contact due to its simplicity and high empirical accuracy, where it can be seen implemented in a wide variety of engineering applications, including, contact in elastic structures, most granular media simulations, and more. 
We therefore build a novel numerical framework based on Coulomb friction.

Aside from friction formulations, contact handling methods can generally be divided into three distinct categories: impulse methods, constraint-based optimization methods, and penalty energy methods.
As the name suggests, impulse methods compute contact forces based on the required impulse to keep rod segments from penetrating, with an example being the impulse force model by Spillmann et al.~\cite{spillmann2008adaptive}. 
Although computationally efficient and straightforward to implement, unrealistic visual jittering often occurs when simulations use sufficiently large time steps as the generated forces are handled explicitly~\cite{choi2021imc}. Therefore, impulse methods often must either deal with insufficient physical accuracy or use sufficiently small time steps.

Constraint-based methods treat frictional contact as a constrained optimization problem. Jean and Moreau~\cite{jean1987dynamics, jean1992unilaterality} implemented convex analysis to propose using unilateral constraints to solve dry friction in granular media. Alart and Curnier.~\cite{alart1991mixed} developed the first approach to solving constraint-based contact dynamics using Newton's method to find the root of a non-smooth function.
In graphics, Daviet et al.~\cite{daviet2011hybrid} combined an analytical solver with the complementary condition from~\cite{alart1991mixed} to capture Coulomb friction in elastic fibers. In Ref.~\cite{kaufman2014adaptive}, the algorithm from~\cite{daviet2011hybrid} was incorporated with a nonlinear elasticity solver to simulate frictional contacts in assemblies of Discrete Elastic Rods~\cite{bergou2008discrete, bergou2010discrete}. Based on previous work, Daviet~\cite{daviet2020simple} proposed a general constraint-based framework for simulating contact in thin nodal objects. 
Overall, constraint-based methods can often produce physically realistic results but are inherently more difficult to implement than impulse and penalty methods (though the growth of open source code has alleviated this considerably).
Arguably the largest drawback of constraint-based methods, additional computational costs are incurred at each solving iteration as frictional contact forces must be introduced as additional degrees of freedom in order to satisfy the complementary condition between frictional contact responses and the status of contact regions. This is in contrast to impulse and penalty methods which can obtain contact responses directly based on just configuration-based degrees of freedom.


The final contact method type, penalty energy methods, utilize a formulated ``contact energy" whose gradient is treated as the contact force. 
Due to the requirement of a smooth differentiable gradient (and Hessian for implicit formulations), such methods utilize smooth differentiable functions to best approximate the behavior of frictional contact \cite{li2020incremental, choi2021imc, patil2020topological}.
These methods have become popular in recent times as they have been shown capable of generating accurate frictional contact~\cite{choi2021imc, patil2020topological} while remaining simple to implement (relative to constraint-based methods) and computationally efficient. 
Building upon this, we propose Implicit Contact Model (IMC), a fully-implicit penalty-based contact model for frictional contact based on our previous work in Ref.~\cite{choi2021imc}.
We improve upon this iteration by (1) reformulating frictional contact to be fully-implicit for enhanced physical accuracy, (2) squaring our contact energy term for more rigid contact, (3) changing our smoothly approximated distance formulation to a more stable piecewise analytical formulation, and (4) adding a line search method for increased robustness.
Our proposed numerical framework can generate contact for any rod-rod contact scenario and can also generate contact for 3D meshes with proper alterations.

In this paper, we choose to showcase our frictional contact model by simulating the novel and difficult contact scenario of flagella bundling~\cite{flores2005study,cisneros2008unexpected,reigh2012synchronization,maniyeri2014numerical,hintsche2017polar,nguyen2018impacts, lee2018bacterial,huang2021numerical,powers2002role}, a significant natural phenomena that occurs when micro-organisms with multiple flagella swim in fluid (e.g., \textit{Escherichia coli} and \textit{Salmonella typhimurium}~\cite{kim2003macroscopic}).
Each flagellum consists of a rotary ``head'', a short flexible hook, and a helical filament.
By rotating their filaments, these micro-organisms can navigate their environments through sophisticated manipulation of the solid-fluid interaction between their flexible structures and the surrounding flows. 
This has led to biomimicry, where flagella have inspired the design of several soft robot locomotion strategies in viscous fluids~\cite{magdanz2013development,ye2013rotating,beyrand2015multi,son2013bacteria, rusconi2014bacterial}. 
However, studying the mechanics of flagella is exceptionally challenging due to flagella possessing radii smaller than their optical wavelengths as well as having high rotational speeds~\cite{jawed2015propulsion}. 

This highlights the necessity for accurate simulators, the development of which is nontrivial due to the multifaceted problem of having to deal with hydrodynamic interactions, geometrically nonlinear deformations, and realistic contact handling. Prior works~\cite{reigh2012synchronization, lee2018bacterial, huang2021numerical} usually consider the contact forces as simple repulsive forces. We instead incorporate our more principled frictional contact framework for the task of flagella bundling.
Extensive validation of our generated frictional contact forces can be seen in Appendix C~\cite{SM}.
We note that despite this, our simulation results have not been validated for the task of flagella bundling in particular.
Still, to the best of our knowledge, this paper is the first to design a fast and visually realistic simulator that can capture the bundling phenomena of multiple flagella rotating in low Reynolds number fluids.
We believe our simulator has promise for aiding the study of flagella bundling where we conduct an informative parametric study for the bundling process using our framework in Appendix D~\cite{SM}.
It is also a first step towards efficient, physically realistic data generators for training bio-inspired flagellar robots for data-driven control approaches.

The primary contributions of our work are outlined below.
\begin{itemize}
	\item We propose a fully-implicit penalty-based frictional contact model that has improved computational efficiency and accuracy compared with our previous work in Ref.~\cite{choi2021imc}.
	\item We formulate a full end-to-end framework for the novel and difficult contact scenario of flagella bundling in low Reynolds number fluids, which incorporates a DDG-based simulation framework, our frictional contact framework, and fluid-solid interaction. 
	\item We conduct an in-depth side-by-side comparison between our proposed method (IMC) with the state-of-the-art (IPC)~\cite{li2020incremental} and show that we are able to achieve faster convergence at the price of losing guarantee of non-penetration.
	\item We show visually convincing results for the sticking slipping phenomena of friction.
\end{itemize}

The remainder of the paper is as follows. In Sec.~\ref{sec:der}, we introduce the DDG-based framework for simulating a flexible rod. Next, in Sec.~\ref{sec:methodology}, we formulate our frictional contact model. Simulation results for IMC are then shown in Sec.~\ref{sec:results} along with side-by-side comparisons with IPC and analysis of friction performance. Finally, in Sec. \ref{sec:conclusion}, we provide concluding remarks as well as potential future research directions. 
Further details on the fluid-solid interaction model, IMC algorithmic components, and miscellaneous information can be found in Supplementary Information Appendices A, B, and E, respectively~\cite{SM}.
All source code used in this paper is released publicly and can be found at \url{https://github.com/StructuresComp/rod-contact-sim}.

\section{Discrete Elastic Rods}
\label{sec:der}

\begin{figure}[h]
    \centering
    \includegraphics[width=0.8\textwidth]{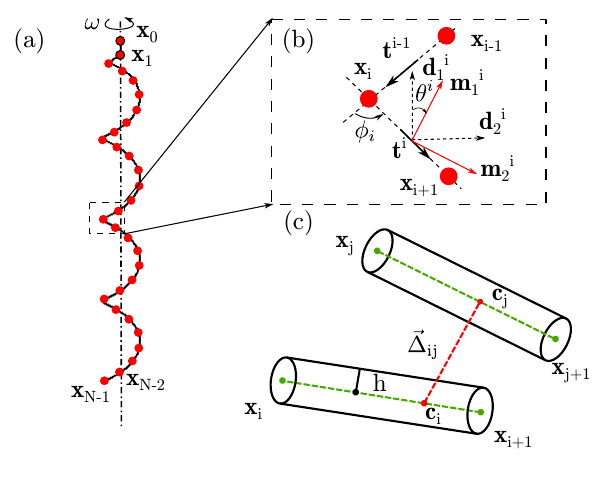}
    \caption{(a) Discrete schematic of a flagellar elastic rod. Nodes $\mathbf x_0$, $\mathbf x_1$, and the edge between $\mathbf x_0$ and $\mathbf x_1$ is clamped along the dashed centerline and rotated with an angular velocity $\omega$. The rest of the nodes constitute the helical flagellum which revolves around the centerline. (b) A zoomed in snapshot of two edges showcasing their reference frame, material frame, turning angles, and twist angles. (c) Illustration of two edges approaching contact. The green dots showcase the nodes of the edges while the green dashed lines denote the centerlines of the edges. The red dashed line denotes the vector $\vec \Delta$ whose norm is the minimum distance $\Delta$ between the edges. $\vec \Delta$ is connected to edges $i$ and $j$ by $\mathbf c_i = \mathbf x_i + \beta_i (\mathbf x_{i+1} - \mathbf x_i)$ and $\mathbf c_j = \mathbf x_j + \beta_j (\mathbf x_{j+1} - \mathbf x_j)$ where $\beta_i, \beta_j \in [0, 1]$.
    As $\Delta$ approaches the contact threshold $2h$, repulsive forces increase at an exponential rate, thus enforcing non-penetration.
    } 
    \label{fig:DER}
\end{figure}

In order to simulate the geometric nonlinear behaviors of flagella in viscous fluids, we utilize the DDG-based framework Discrete Elastic Rods (DER)~\cite{bergou2008discrete, bergou2010discrete}.
As shown in Fig.~\ref{fig:DER}(a), DER expresses the centerline of an elastic rod with $N$ discrete nodes: $\mathbf x_0, \mathbf x_1, ... \mathbf x_{N-2}, \mathbf x_{N-1}$. This results in a total of $N-1$ edges where $\mathbf e^i = \mathbf x_{i+1} - \mathbf x_i$.
Note that for DER, we use subscripts to denote indices for quantities associated with nodes and superscripts for indices for quantities associated with edges. 
Following this, each edge $\mathbf{e^i}$ is described using two orthogonal frames: a reference frame $\{ \mathbf t^i, \mathbf d^i_1, \mathbf d^i_2\}$ and a material frame $\{ \mathbf t^i, \mathbf m^i_1, \mathbf m^i_2\}$ as shown in Fig.~\ref{fig:DER}(b).
The reference frame is predefined at initial time $t = 0$s. The material frame shares the same director $\mathbf t^i = \mathbf e^i / \lVert \mathbf e^i \rVert$ as the reference frame and is obtainable through a twist angle $\theta^i$ with respect to the reference frame. 
A total of $N$ nodes, each represented by a Cartesian coordinate $\mathbf x_i \in \mathbb{R}^3$, and $N-1$ twist angles constitute a total of $4N - 1$ degrees of freedom: $\mathbf q = [\mathbf x_0, \theta^0, \mathbf x_1, ..., \mathbf x_{N-2}, \theta^{N-2}, \mathbf x_{N-1}]$.

To simulate the elastic properties of a rod, we must compute elastic energies as defined by strain. Based off of Kirchhoff's rod model \cite{kirchhoff1859uber}, strains can be divided into three categories: stretching, bending, and twisting. Starting off, the stretching strain of an edge $\mathbf e^i$ is described by
\begin{equation}
\label{eq:stretchStrain}
    \epsilon^i = \frac{\lVert \mathbf e^i \rVert}{\lVert \bar{\mathbf e}^i \rVert} - 1,
\end{equation}
where $\lVert \bar{\mathbf e}^i \rVert$ is the undeformed length of edge $\mathbf{e}^i$. From hereafter, quantities with a $\ \bar{} \ $ represent their values in their undeformed state.

Moving forwards, the bending strain for a node $\mathbf x_i$ is evaluated by a curvature binormal which captures the misalignment between two consecutive edges:
\begin{equation}
\label{eq:curvatureBinormal}
    (\kappa \mathbf b)_i = \frac{2 \mathbf e^{i-1} \times \mathbf e^i}{\lVert \mathbf e^{i-1}\rVert \ \lVert \mathbf e^i \rVert + \mathbf e^{i-1} \cdot \mathbf e^i }.
\end{equation}
Here, $\lVert (\kappa \mathbf b)_i \rVert = 2 \tan(\phi_i/2)$, where $\phi_i$ is the turning angle shown in Fig.~\ref{fig:DER}(b). The material curvatures are the components of the curvature binormal $(\kappa \mathbf b)_i$ via the directors of the material frame:
\begin{equation}
\label{eq:bendingStrain}
\begin{aligned}
\kappa_i^{(1)} &= \frac{1}{2}(\mathbf m^{i-1}_2 + \mathbf m^{i}_2) \cdot (\kappa \mathbf b)_i,\\
\kappa_i^{(2)} &= \frac{1}{2}(\mathbf m^{i-1}_1 + \mathbf m^{i}_1) \cdot (\kappa \mathbf b)_i.
\end{aligned}
\end{equation}

Finally, the twisting strain for a node $\mathbf x_i$ is computed as
\begin{equation}
\label{eq:twistingStrain}
\tau_i = \theta^i - \theta^{i-1} + m^i_{\textrm{ref}},
\end{equation}
where $m^i_{\textrm{ref}}$ is the difference between the two consecutive reference frames of the $i$ and $i-1$-th edges. 

With all strains defined, we can now formulate the stretching, bending, and twisting energy of a discretized elastic rod:
\begin{equation}
\label{eq:elasticEnergies}
\begin{aligned}
E_s & = \frac{1}{2} \sum_{i=0}^{N} EA (\epsilon^i)^2 \lVert \bar{\mathbf e}^i \rVert, \\
E_b & = \frac{1}{2} \sum_{i=1}^{N} \frac{1}{\Delta l_i} \left[ EI_1\left(\kappa_i^{(1)} - \bar{\kappa}_i^{(1)}\right)^2  + EI_2\left(\kappa_i^{(2)} - \bar{\kappa}_i^{(2)}\right)^2 \right], \\
E_t & = \frac{1}{2} \sum_{i=1}^{N} \frac{GJ}{\Delta l_i} (\tau_i)^2,
\end{aligned}
\end{equation}
where $E$ is Young's Modulus; $A$ is the cross-sectional area; $G$ is the shear modulus; $J$ is the polar second moment of area along tangent $\mathbf t^i$; $I_1$ and $I_2$ are moments of inertia along material directors $\mathbf m_1^i$ and $\mathbf m_2^i$; and $\Delta l_i = (\lVert e^{i} \rVert + \lVert e^{i+1} \rVert)/2$ is the Voronoi length.

Next, the internal forces (for each nodal degree of freedom) and moments (for each twist degree of freedom) can be obtained via the partial derivative of the sum of elastic energies given in Eq.~\ref{eq:elasticEnergies} as shown:
\begin{equation}
\label{eq:DER_Fint}
F^{\textrm{int}}_i =  - \frac{\partial (E_s + E_b + E_t)}{\partial {q}_i} \  \forall \ i \in [0, 4N-1].
\end{equation}
These quantities result in a $4N-1$ sized force vector $\mathbf F^\textrm{int}$. Following this, we can write the system of equations of motions as the sum of inertial terms, internal forces, and external forces (e.g. contact, friction, gravity). This results in the equation
\begin{equation}
\label{eq:DER_soe}
\mathbf F   \equiv  \mathbb M \ddot{\mathbf q} - \mathbf F^{\textrm{int}} - \mathbf F^{\textrm{ext}} = 0,
\end{equation}
where $\mathbb M$ is the diagonal mass matrix, $\ddot{\mathbf q}$ is the second derivative of the DOFs with respect to time, $\mathbf F^\textrm{ext}$ is the external force vector, and $\mathbf F$ is the total force. This external force vector will be made up of our contact forces described in Sec.~\ref{sec:methodology} as well as viscous drag forces from solid-fluid interactions described in Supplementary Information~\cite{SM}.
As Eq. \ref{eq:DER_soe} is a root-finding problem, we use Newton's method to solve the system of equations to march through time.
Note that the Jacobian of the viscous drag forces is unobtainable. Therefore, viscous drag forces are treated explicitly, i.e. their Jacobian is simply ignored.
Still, as our contact model is fully implicit, we are able to robustly simulate flagella bundling regardless of the explicitly added forces, as we will soon show.


\section{Contact Model Methodology}
\label{sec:methodology}
Moving forwards, we denote the following vector concatenation describing an edge-to-edge contact pair: $\mathbf x_{ij}:=(\mathbf x_i, \mathbf x_{i+1}, \mathbf x_j, \mathbf x_{j+1}) \in \mathbb R^{12}$, where $|j-i| > 1$ to exclude consecutive edges from consideration when enforcing contact. We describe the set of all valid edge combinations as $\mathcal X$. In future equations, we simply denote the subscriptless $\mathbf x$ as an arbitrary edge combination for clarity. 
We design a contact energy $E(\Delta(\mathbf x))$ to increase as the minimum distance $\Delta$ between two bodies approaches a contact threshold ($2h$ for our application where $h$ is the radius of the flagella).
With this, the contact energy gradient $-k\nabla_{\mathbf x} E(\mathbf x) \in \mathbb R^{12}$ is used as the contact forces while the contact energy Hessian $-k\nabla^2_{\mathbf x} E(\mathbf x) \in \mathbb R^{12 \times 12}$ is used as the contact force Jacobian, where $k$ is the contact stiffness which scales the contact forces appropriately to enforce non-penetration. 
In the upcoming sections, we will now formulate contact energy $E(\Delta(\mathbf x))$, minimum distance between two edges $\Delta(\mathbf x)$, as well as friction. 

\begin{figure}[!t]
    \centering
    \includegraphics[width=\textwidth]{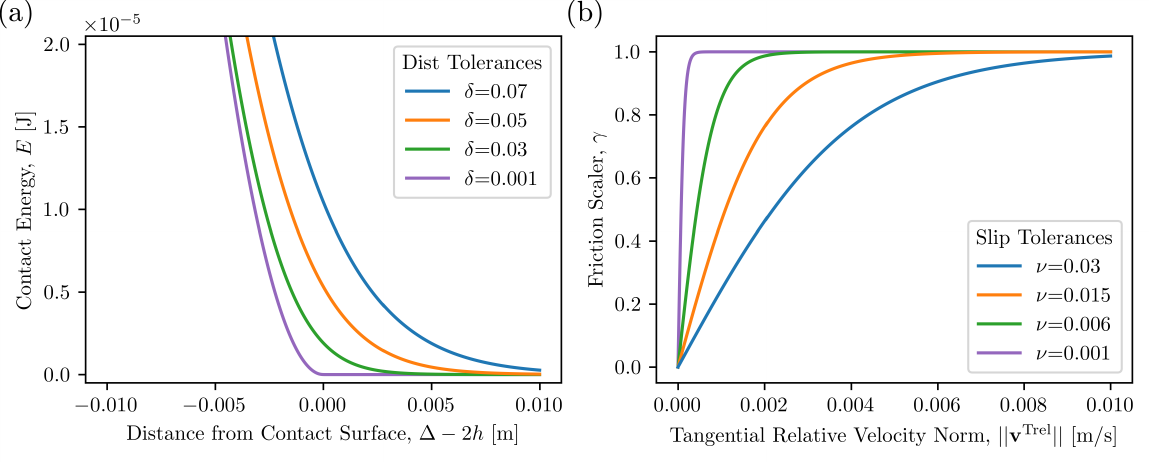}
    \caption{Plots for the approximation functions in (a) Eq. \ref{eq:contact_energy} and (b) Eq. \ref{eq:gamma} with varying tolerance values. Note that some of the tolerances displayed are unrealistically large for clarity.}
    \label{fig:approx_funcs}
\end{figure}

\subsection{Contact Energy}
\label{subsec:contact_energy}
In the ideal setting, contact energy must satisfy two properties: (1) it is zero for any distance $\Delta > 2h$ and (2) it is non-zero at exactly distance $\Delta = 2h$. A Heaviside step function can essentially describe these properties. Such a function is non-smooth with a very sudden discontinuous change in value, and therefore, cannot be solved reliably by root-finding algorithms such as Newton's method. 
To remedy this, IPC uses the following energy formulation to smoothly approximate contact:
\begin{equation}
\label{eq:ipc_contact_energy}
    E^\textrm{IPC}(\Delta, \delta) = \begin{cases}
                                -(\Delta - (2h + \delta))^2 \ln \left(\frac{\Delta}{2h + \delta}\right), & \Delta \in (2h, 2h + \delta) \\
                                0 & \Delta \geq 2h + \delta,
                                 \end{cases}
\end{equation}
where $\delta$ is the distance tolerance that defines the region $(2h, 2h+\delta)$ for which non-zero forces are experienced.
This contact energy approaches $\infty$ when $\Delta$ decreasingly approaches $2h$ and is therefore undefined for the region $\Delta \leq 2h$. Although this barrier formulation allows IPC to strictly enforce non-penetration, the solver must be careful never to allow any contact pairs in the penetration zone and/or venture into this undefined region during the optimization process. This is ensured by the inclusion of a custom line search method which conservatively sets an upper bound for the Newton update coefficient $\alpha$. 

In contrast to this, we design our energy formulation to allow for optimization into the penetrated region, thus expanding the range contact forces are experienced from $\Delta \in (2h, 2h + \delta)$ to $\Delta \in (0, 2h + \delta)$. This in turn allows us to take advantage of more aggressive line search methods, which leads to faster convergence for the flagella contact problem. 
Although this in theory allows our model to be susceptible to penetration, a sufficient contact stiffness $k$ remedies this issue. 
We provide a method that adaptively sets an appropriate stiffness value in Supplementary Information~\cite{SM}. 
In addition, to further ensure non-penetration, we take our previous energy formulation from \cite{choi2021imc} and square it so that our gradient grows exponentially instead of linearly.
In the end, we use the smooth approximation
\begin{equation}
\label{eq:contact_energy}
    E(\Delta, \delta) = \begin{cases}
                                (2h - \Delta)^2 & \Delta \in (0, 2h-\delta] \\
                                \left(\frac{1}{K_1} \log(1 + \exp(K_1 (2h - \Delta)))\right)^2 & \Delta \in (2h - \delta, 2h + \delta) \\
                                0 &  \Delta \geq 2h + \delta,
                                 \end{cases}
\end{equation}
where $K_1 = 15 / \delta$ indicates the stiffness of the energy curve. 

We incorporate the piecewise term $(2h - \Delta)^2$ for two reasons. First, this term equivalently models our energy formulation for the region $\Delta \leq 2h-\delta$ and has a simpler gradient and Hessian, resulting in computational efficiency. Second, and more importantly, the piecewise term also ensures numeric stability by preventing the exponential term in Eq. \ref{eq:contact_energy} from exploding.
We show our plotted energy term in Fig. \ref{fig:approx_funcs}(a) for various $\delta$ values.
As shown, the energy starts to increase at an exponential rate as $\Delta$ decreases towards the contact limit which is shown as 0 here. As $\delta$ decreases, more realistic contact is achieved (enhancing accuracy) in exchange for a stiffer equation (more difficult to converge).

\subsection{Computing Distance}
\label{subsec:distance}

As mentioned in \cite{li2020incremental}, the minimum distance between two edges $(\mathbf x_i, \mathbf x_{i+1})$ and $(\mathbf x_j, \mathbf x_{j+1})$ can be formulated as the constrained optimization problem
\begin{equation}
\label{eq:min_dist_opt}
    \Delta = \min_{\beta_i, \beta_j} || \mathbf x_i + \beta_i (\mathbf x_{i+1} - \mathbf x_i) - (\mathbf x_j + \beta_j (\mathbf x_{j+1} - \mathbf x_j)) || \ni 0 \leq \beta_i, \beta_j \leq 1,
\end{equation}
where $\beta_i$ and $\beta_j$ represent the contact point ratios along the respective edges.
Minimum distance between two edges can be classified into three distinct categories: point-to-point, point-to-edge, and edge-to-edge. As the names suggest, these classifications depend on which points of the edges the minimum distance vector $\vec \Delta$ lies as described by $\beta_i$ and $\beta_j$ shown in Fig. \ref{fig:DER}(c).

In our previous work \cite{choi2021imc}, we altered Lumelsky's edge-to-edge minimum distance algorithm \cite{lumelskey1985min_dist} (which implicitly computes the $\beta$ values) to be fully differentiable through smooth approximations. In this work, we now change the distance formulation to use piecewise analytical functions as shown below in Eqs. \ref{eq:dist_p2p}, \ref{eq:dist_p2e}, and \ref{eq:dist_e2e}, similar to \cite{li2020incremental}, as we found more stable performance compared to our smooth formulation despite the non-smooth Hessian when changing between contact categories. 

We now describe the conditions for each contact type classification.
First, if $\vec \Delta$ lies on the ends of both edges (i.e. both $\beta$ constraints are active), then the distance formulation degenerates to the point-to-point case which can easily be solved using the Euclidean distance formula, 
\begin{equation}
\label{eq:dist_p2p}
    \Delta^{PP} = || \mathbf x_a - \mathbf x_b ||,
\end{equation}
where $\mathbf x_a$ and $\mathbf x_b$ are the nodes for first and second edges in contact, respectively.

If $\vec \Delta$ only lies on one end of one rod (i.e. only one $\beta$ constraint is active), then the contact type degenerates to point-to-edge. This can be solved as
\begin{equation}
\label{eq:dist_p2e}
        \Delta^{PE} = \frac{|| (\mathbf x_a - \mathbf x_b) \times (\mathbf x_b - \mathbf x_c)||}{|| \mathbf x_a - \mathbf x_b ||},
\end{equation}
where $\mathbf x_a$ and $\mathbf x_b$ are the nodes of the edge for which the minimum distance vector does not lie on an end and $\mathbf x_c$ is the node of the edge which the minimum distance vector does lie on.
Finally, edge-to-edge distance (i.e. no active constraints) for the $i$-th and $j$-th edges can be solved as
\begin{equation}
\label{eq:dist_e2e}
\begin{aligned}
        \mathbf u &= (\mathbf x_{i+1} - \mathbf x_{i}) \times (\mathbf x_{j+1} - \mathbf x_{j}), \\
        \Delta^{EE} &= | (\mathbf x_{i} - \mathbf x_{j}) \cdot \hat{\mathbf u} |,
\end{aligned}
\end{equation}
where $\ \hat{} \ $ indicates a unit vector. With $\Delta$ fully defined, this concludes our contact energy formulation. To correctly classify contact pairs, we use Lumelsky's algorithm to compute $\beta$ values.

\begin{algorithm}[!t]
\SetAlgoLined
\LinesNumbered
\DontPrintSemicolon
\KwIn{$ \mathbf{x}, \mathbf x_0, k, \delta, \nu$}
\KwOut{$ \mathbf{F}^\textrm{c}, \mathbf{J}^\textrm{c}, \mathbf{F}^\textrm{fr}, \mathbf{J}^\textrm{fr}$}
\SetKwFunction{FMain}{IMC}
\SetKwProg{Fn}{Function}{:}{}
\SetKwFunction{genContact}{genContact}
\SetKwFunction{genFriction}{genFriction}
\SetKwFunction{genFrictionPartials}{genFrictionPartials}
\SetKwFunction{collisionDetection}{collisionDetection}
\SetKwFunction{updateConStiffness}{updateConStiffness}
\SetKwFunction{newtonDamper}{newtonDamper}
    \Fn{\FMain{$\mathbf{x}, \mathbf x_0, k, \delta, \nu$}}
    {
        $\mathbf v \gets \mathbf x - \mathbf x_0$ \tcp*{compute velocity}
        $\mathbf{F}^\textrm{c}, \mathbf J^\textrm{c} \gets \genContact(\mathbf x, \delta)$ \tcp*{Eq.\ref{eq:contact_energy}}
        $\mathbf{F}^\textrm{c} \gets k \mathbf{F}^\textrm{c}$ \tcp*{scale by contact stiffness}
        $\mathbf J^\textrm{c} \gets k \mathbf J^\textrm{c}$\tcp*{$\mathbf J^\textrm{c} \equiv \nabla_{\mathbf x} \mathbf F^\textrm{c}$}
        $\mathbf{F}^\textrm{fr} \gets \genFriction(\mathbf x, \mathbf{v}, \mathbf{F}^\textrm{c}, \nu)$ \tcp*{Eq.\ref{eq:friction}}
        $\nabla_{\mathbf x} f, \nabla_{\mathbf F^\textrm{c}} f \gets \genFrictionPartials(\mathbf x, \mathbf{v}, \mathbf{F}^\textrm{c}, \nu)$ \tcp*{$\mathbf F^\textrm{fr} \equiv f(\mathbf x, \mathbf F^\textrm{c})$}
        $\mathbf J^\textrm{fr} \gets \nabla_{\mathbf x} f + \nabla_{\mathbf F^\textrm{c}} f  \nabla_{\mathbf x} \mathbf F^{\textrm{c}}$  \tcp*{Eq.\ref{eq:fric_chain_rule}}
        \Return $\mathbf{F}^\textrm{c}, \mathbf{J}^\textrm{c}, \mathbf{F}^\textrm{fr}, \mathbf{J}^\textrm{fr}$ 
    }
\caption{Implicit Contact Model}
\label{alg:IMC}
\end{algorithm}

\subsection{Adding Friction} 
\label{subsec:friction}
Similar to before, we model friction according to Coulomb's friction law, which describes the conditions necessary for two solids to transition between sticking and sliding. This law states that the frictional force $F^\textrm{fr}$ is (1) equal to $\mu F^\textrm{n}$ during sliding, (2) is in the region of $[0, \mu F^\textrm{n})$ when sticking, and (3) is independent of the magnitude of velocity.
Here, $\mu$ is the friction coefficient and $F^\textrm{n}$ is the normal force experienced by the body. 

Let us denote the following equivalencies for clarity: $\mathbf F^\textrm{c} \equiv k\nabla_\mathbf{x} E$ and $\mathbf J^\textrm{c} \equiv k\nabla^2_\mathbf{x} E $.
Following this, for a contact pair $\mathbf x_{ij} := (\mathbf x_i, \mathbf x_{i+1}, \mathbf x_j, \mathbf x_{j+1})$, we can obtain the normal force on the $i$-th and $i+1$-th nodes as $F^\textrm{n}_i = \lVert \mathbf F^\textrm{c}_i \rVert$ and $F^\textrm{n}_{i+1} = \lVert \mathbf F^\textrm{c}_{i+1} \rVert$, respectively. This in turn allows us to obtain the contact norm vector
\begin{equation}
    \begin{aligned}
        \mathbf n_i &= \frac{ \mathbf F^\textrm{c}_{i} + \mathbf F^\textrm{c}_{i+1}}{\lVert \mathbf F^\textrm{c}_{i} + \mathbf F^\textrm{c}_{i+1}\rVert}.
    \end{aligned}
\label{eq:contact_norm}
\end{equation}
The direction of friction is then the tangential relative velocity between edges $i$ and $j$. 
To compute this, we must first compute the relative velocities of the edges at the point of contact, which can be done using $\beta_i, \beta_j \in [0, 1]$ as shown below:
\begin{equation}
\label{eq:relative_velo}
    \begin{aligned}
        \mathbf v^\textrm{e}_i &= (1-\beta_i) \mathbf v_i + \beta_i \mathbf v_{i+1},\\
        \mathbf v^\textrm{e}_j &= (1 - \beta_j) \mathbf v_j + \beta_j \mathbf v_{j+1},\\
        \mathbf v^\textrm{rel} &= \mathbf v^\textrm{e}_i - \mathbf v^\textrm{e}_j, 
    \end{aligned}
\end{equation}
where $\mathbf v_i, \mathbf v_{i+1}, \mathbf v_j$, and $\mathbf v_{j+1}$ are the velocities of the $i$-th, $i+1$-th, $j$-th, and $j+1$-th nodes, respectively.
The tangential relative velocity of edge $i$ with respect to edge $j$ can then be computed as
\begin{equation}
    \begin{aligned}
        \mathbf v^\textrm{Trel} &= \mathbf v^\textrm{rel} - (\mathbf v^\textrm{rel} \cdot \mathbf n_i) \mathbf n_i, 
    \end{aligned}
\end{equation}
where $\hat{\mathbf v}^\textrm{Trel} = \mathbf v^\textrm{Trel} / ||\mathbf v^\textrm{Trel}||$ is our friction direction.

Now, we must also make our contact model capable of simulating the transition between sticking and sliding. Coulomb's law tells us that $\lVert \mathbf v^\textrm{Trel}\rVert=0$ during static friction and that $\lVert \mathbf v^\textrm{Trel}\rVert>0$ for sliding friction. Sticking occurs up until the tangential force threshold $\mu F^\textrm{n}$ is surpassed, after which sliding begins. This relation (similar to ideal contact energy) can also be described by a modified Heaviside step function. For the same reasons as before, we replace this step function for another smooth approximation described by 
\begin{equation}
\label{eq:gamma}
        \gamma \left(\lVert \mathbf v^\textrm{Trel} \rVert, \nu \right) = \frac{2}{1 + \exp\left(-K_2 \lVert \mathbf v^\textrm{Trel} \rVert \right)} - 1,
\end{equation}
where $\nu$ (m/s) is our desired slipping tolerance and $K_2(\nu) = 15 / \nu$ is the stiffness parameter. 
As shown in Fig. \ref{fig:approx_funcs}(b), $\gamma \in [0, 1]$ smoothly scales the friction force magnitude from zero to one as $\lVert \mathbf v^\textrm{Trel} \rVert$ increases from zero.
The slipping tolerance describes the range of velocities $(0, \nu)$ for which a friction force $< \mu F^\textrm{n}$ is experienced. In other words, we consider velocities within this range to be ``sticking".

Finally, the friction experienced by a node $i$ for a contact pair $\mathbf x_{ij}$ can be described as

\begin{equation}
    \mathbf F^\textrm{fr}_i = - \mu \gamma \hat{\mathbf v}^\textrm{Trel} F^\textrm{n}_i.
\label{eq:friction}
\end{equation}

With friction fully defined, we can now formulate the friction Jacobian $\nabla_\mathbf{x} \mathbf F^\textrm{fr}$.
Note that due to Eq. \ref{eq:relative_velo}, our formulation depends on $\beta(\mathbf x)$, which means that the gradient $\nabla_\mathbf{x} \beta$ is required. 
We can avoid this computation through the realization that the magnitudes of the contact forces $\mathbf F^\textrm{c}_i$ and $\mathbf F^\textrm{c}_{i+1}$ have an underlying linear relationship with $\beta$ where
\begin{equation}
    \begin{aligned}
        \mathbf F^\textrm{c}_i &= (1 - \beta) (\mathbf F^\textrm{c}_i + \mathbf F^\textrm{c}_{i+1}), \\
        \mathbf F^\textrm{c}_{i+1} &= \beta (\mathbf F^\textrm{c}_i + \mathbf F^\textrm{c}_{i+1}).
    \end{aligned}
\end{equation}
Therefore, we can obtain $\beta$ by simply solving 
\begin{equation}
    \begin{aligned}
        \beta =  \frac{\lVert \mathbf F^\textrm{c}_{i+1} \rVert}{\lVert \mathbf F^\textrm{c}_i + \mathbf F^\textrm{c}_{i+1} \rVert}.
    \end{aligned}
\end{equation}
We can now treat $\beta$ as a function of $\mathbf F^\textrm{c}$, resulting in a simplified chain ruling procedure. Let us denote Eq. \ref{eq:friction} as the functional $f(\mathbf x, \mathbf F^\textrm{c}(\mathbf x)$). The friction Jacobian can then be computed through chain rule as
\begin{equation}
    \nabla_{\mathbf x} \mathbf F^\textrm{fr} = \nabla_{\mathbf x} f + \nabla_{\mathbf F^\textrm{c}} f \nabla_{\mathbf x} \mathbf F^{\textrm{c}}.
\label{eq:fric_chain_rule}
\end{equation}
This concludes our fully implicit friction scheme. Full psuedocode for the IMC algorithm can be found in Alg. \ref{alg:IMC}. 

\section{Simulation Results}
\label{sec:results}
In this section, we showcase extensive quantitative and qualitative results for IMC.
First, we discuss all our simulation parameters.
We then conduct a detailed comparison between IMC and the state-of-the-art contact handling method: Incremental Potential Contact (IPC) \cite{li2020incremental}. 
Afterwards, we showcase comprehensive results concerning friction and display IMC's ability to simulate the sticking sliding transition.

\begin{figure}[!t]
    \centering
    \includegraphics[width=\textwidth]{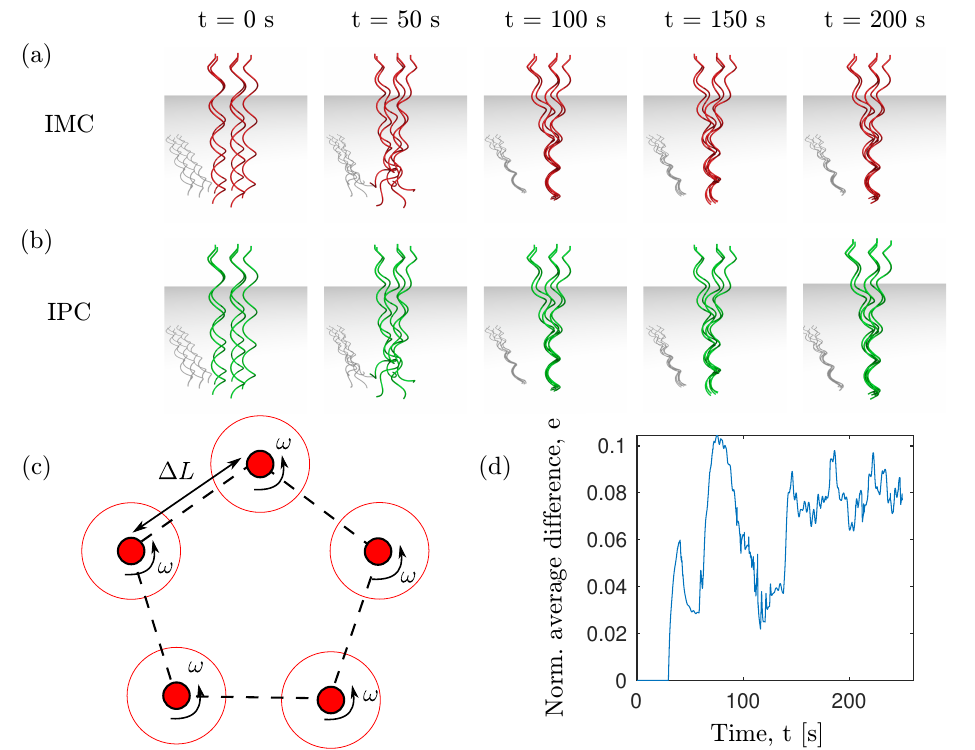}
    \caption{Rendered snapshots for $M=5$ flagella simulated by (a) IMC and (b) IPC. We can observe that there is great qualitative agreement between both methods at the shown time steps. (c) A top down visualization of boundary conditions applied to the highest nodes (filled in red circles) of each flagella as well as the angular rotation $\omega$ applied to them. The larger hollow red circles represent the rest of the helical flagella. (d) The norm of the average difference in the nodal positions for the flagella simulated by IMC and IPC with respect to time.}
    \label{fig:imc_vs_ipc}
\end{figure}

\begin{table}[!t]
\renewcommand{\arraystretch}{1.1}
\caption{IMC vs. IPC \cite{li2020incremental} run time data. Simulations are run for a total of 250 seconds with a time step size of $\Delta t = 1$ ms and a rotation speed of $\omega = 15$ rad/s.
The contact model used can be seen in the far left column. Next to this, $M$ indicates the number of flagella. AIPTS stands for average iterations per time step. ATPTS stands for average time per time step. Total Iters indicates the total number of Newton's iterations that were necessary to complete the simulation. The Total Run Time is the total computational time to completion. Finally, RTI stands for run time improvement and is the ratio of improvement between IMC's and IPC's total run time. \newline \newline}
\centering
\begin{tabular}{|c||c|c|c|c|c|c|}
\hline
Model & $M$ & \textbf{AIPTS}	& \textbf{ATPTS} [ms] & \textbf{Total Iters} & \textbf{Total Run Time} [hr] & \textbf{RTI}\\
\hline
\hline
\multirow{4}{4em}{\centering \textbf{IMC}} & 2	& 3.00 & 10.2 & $6.01 \times 10^5$ & 0.57 & 1.82\\
                                & 3 & 3.01 & 21.3 & $6.04 \times 10^5$ & 1.19 & 1.82\\
                                & 5 & 3.02 & 67.5 & $5.39 \times 10^5$ & 3.34 & 1.40\\
                                & 10 & 3.12 & 389.4 & $6.56 \times 10^5$ & 22.77 & 1.22\\
\hline
\multirow{4}{4em}{\centering \textbf{IPC}} & 2	& 4.00 & 18.75 & $7.98 \times 10^5$ & 1.04 & N/A\\
                                & 3 & 4.00 & 39.5 & $7.93 \times 10^5$ & 2.17 & N/A\\
                                & 5 & 4.01 & 95.3 & $7.09 \times 10^5$ & 4.68 & N/A\\
                                & 10 & 4.02 & 477.47 & $8.45 \times 10^5$ & 27.88 & N/A\\
\hline
\end{tabular}
\label{tab:metrics}
\end{table}

\subsection{Parameters and Setup}
In the simulation, we design the flagella as right-handed helical rods manufactured with linear elastic material. We set the material properties as follows: Young's modulus was set to $E = 3.00$ MPa; Poisson's ratio was set to 0.5; density of the rod was set to $\rho = 1000$ kg/m$^3$; the cross-sectional radius was set to $h = 1$ mm, and the fluid viscosity was set to 0.1 Pa$\cdot$s. 
Here, a Poisson's ratio of 0.5 was chosen to enforce the flagella to be an incompressible material.
The topologies of the flagella are helices with helical radius $a = 0.01$ m, helical pitch $\lambda = 0.05$ m, and axial length $z_0 = 0.2$ m. These parameters were chosen as they best mimic the geometries of biological flagella found in nature ~\cite{jawed2015propulsion,jawed2017dynamics,rodenborn2013propulsion,huang2021numerical}. 

We explore the bundling phenomena with $M$ flagella ($M = [2, 3, 5, 10]$) where the rotating ends of each flagella is fixed along the $z$-axis as shown in Fig. \ref{fig:DER}(a). These rotating ends are treated as boundary conditions and are spaced out equidistantly so as to form a regular polygon with $M$ angles with side length $\Delta L = 0.03$ m as shown in Fig.~\ref{fig:imc_vs_ipc}(c). 
We set the rotation speed of the flagella ends to $\omega = 15$ rad/s which keeps the Reynolds number in our numerical simulation to be always smaller than $4 \times 10^{-2}$, thus satisfying the Stokes flow. 

Finally, we discretize each flagella into 68 nodes for a total of 67 edges. We found this discretization to have the best trade-off between computational efficiency and accuracy. Furthermore, we set the time step size to $\Delta t = 1$ ms. As the forces generated from our fluid model are handled explicitly, we found 1 ms to be the largest stable time step size before convergence performance became hampered.
A distance tolerance of $\delta = 1 \times 10^{-5}$ was used for all simulations.

\subsection{Comparison between IMC and IPC}
\label{subsec:imc_vs_ipc}

Both IMC and IPC were used to simulate 250 seconds of rotation for scenarios with 2, 3, 5, and 10 flagella as shown in Fig. \ref{fig:flagella_frames}. As the friction coefficient between structures is usually trivial in viscous fluids, we consider purely contact without friction ($\mu = 0$).
First, a side-by-side visual comparison for $M=5$ is shown for IMC and IPC in Fig. \ref{fig:imc_vs_ipc}(a). For various time steps, we can see that the configurations of the flagella are near identical, indicating that both methods have comparable performance.
To further study this similarity, we define normalized average difference $\bar {e}$ to measure the difference in flagella nodal configurations between IMC and IPC:
\begin{equation}
    \bar{e} = \frac{1}{MNh} \sum_{i=0}^{M-1}\sum_{j=0}^{N-1} \left\lVert \mathbf x_{j}^{i, \textrm{IMC}} - \mathbf x_{j}^{i, \textrm{IPC}} \right\rVert,
\end{equation}
The relationship between normalized average difference $\bar e$ and time $t$ is shown in Fig.~\ref{fig:imc_vs_ipc}(d). Here, we can find that the difference between the configurations is quite minimal, further cementing the notion that IMC has comparable performance to IPC despite the loss of non-penetration guarantee.

Where IMC starts to improve upon IPC is in terms of computational efficiency.
Detailed metrics for all runs can be seen in Table \ref{tab:metrics} which showcase the average iteration per time step (AIPTS), average time per time step (ATPTS), total iterations, and total run time. All metrics were recorded using time steps with at least one contact.
Here, we can see that IMC was able to converge with less average iterations than IPC for all flagella cases resulting in significant reductions in total run time. These run time improvements are most drastic for $M=2$ and $M=3$ and start to decrease as $M$ increases further as the RSS force computation starts to become a bottleneck. Regardless, a clear monotonic decrease can still be seen. 


\begin{figure}[!t]
    \centering
    \includegraphics[width=\textwidth]{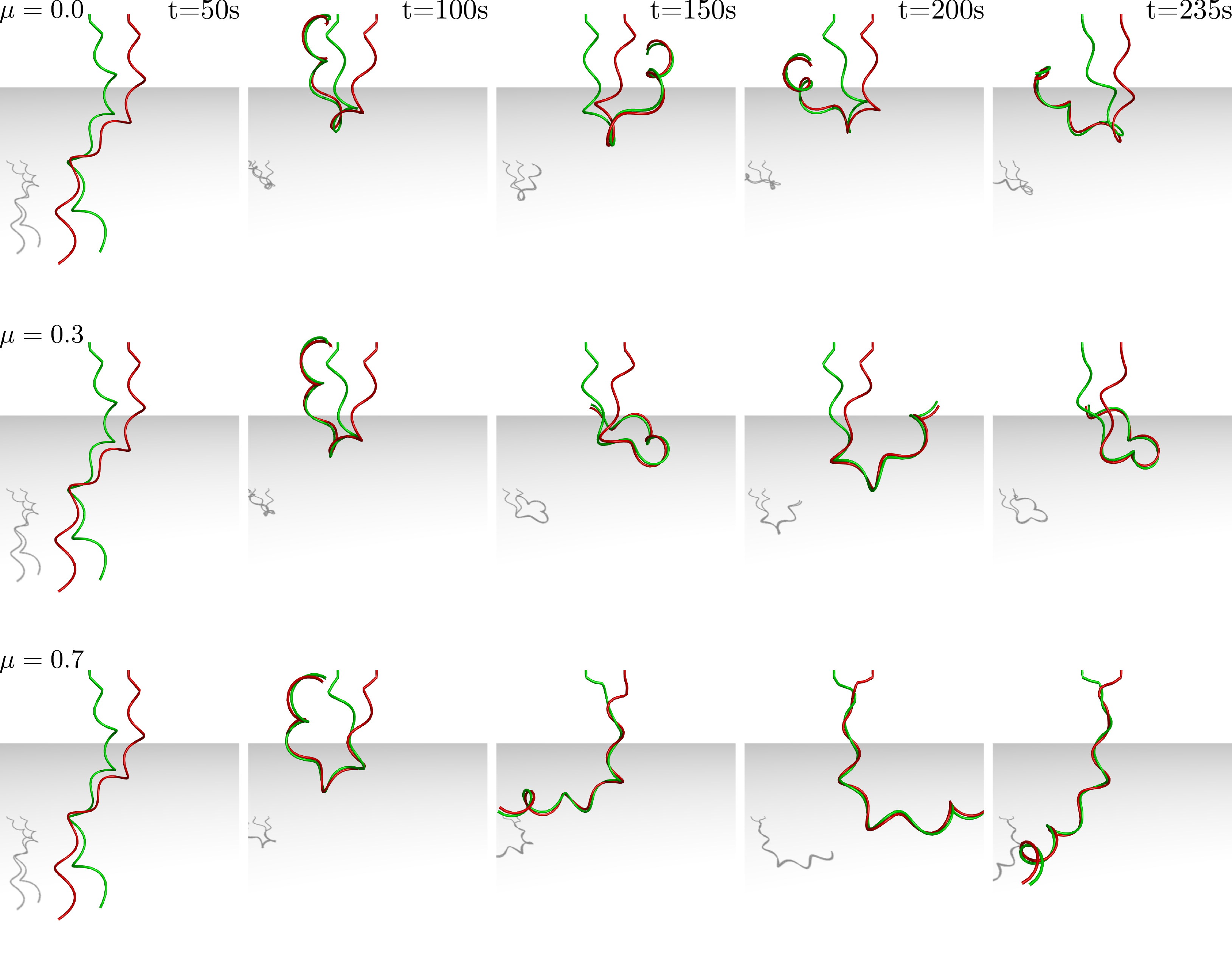}
    \caption{Rendered snapshots for $M=2$ with varying friction coefficients. Each column indicates a moment in time as indicated by the time stamp in the top row. The first row shows the frictionless case $\mu=0$ as a baseline. The second row has $\mu=0.3$ where minor sticking can be observed as the point at where the flagella no longer contact is higher than the frictionless case. Still, $\mu=0.3$ still has plenty of slipping allowing the flagella to not become coiled. As we increase $\mu$ to 0.7 in the third row, we can see the amount of sticking increase, ultimately resulting in the flagella becoming completing coiled.}
    \label{fig:friction_frames}
\end{figure}

\subsection{Friction Example}
\label{subsec:fric_example}
Although friction is usually negligible in a viscous fluid medium, influence of friction on flagella bundling is still intriguing since the effect of friction can become significant as the environment changes (e.g., granular medium). We assume an imaginary viscous environment where the friction coefficient between structures is non-negligible.
We present simulation data for two flagella ($M = 2$) with friction coefficients $\mu = [0.1, 0.2, ..., 1.0]$. 
For all simulations, a slipping tolerance of $\nu=1 \times 10^{-4}$ was used.
All other parameters are kept the same as before. 

We first showcase the sticking slipping phenomena with snapshots for $\mu=[0, 0.3, 0.7]$ in Fig. \ref{fig:friction_frames}.
Intuitively, as $\mu$ increases, we also see the amount of sticking increase as well. Convergence results for all friction examples can be seen in Table \ref{tab:friction_metrics} where average iterations per time step and simulation length are reported. 
Here, we notice two trends. First, for $\mu \geq 0.7$, the time at which the simulation ends starts to decrease from 250 seconds. 
This is because $\mu=0.7$ is the point at which the flagella become completely tangled as shown in the bottom right frame of Fig. \ref{fig:friction_frames}.
As $\mu$ increases past 0.7, the tangling happens earlier and earlier as shown.
Furthermore, we observe that the number of average iterations starts to increase as $\mu$ increases. This is in line with our expectations as larger $\mu$ values result in greater sticking.

\begin{table}[!t]
\setlength{\tabcolsep}{25pt}
\renewcommand{\arraystretch}{1.1}
\caption{Friction results for varying friction coefficients. AIPTS stands for average iterations per time step. Total Iterations indicate the total number of Newton's method iterations that were necessary to complete the simulation. Sim End indicates the total \textit{simulated} time. All simulations were set to run for 250 seconds. As can be seen, simulations with $\mu \geq 0.7$ end earlier due to excessive tangling of the flagella. \newline \newline}
\centering
\begin{tabular}{|c||c|c|c|}
\hline
$\mu$ & \textbf{AIPTS} & \textbf{Total Iterations} & \textbf {Sim End} [sim s] \\
\hline
0.1  & 3.01  & $6.02 \times 10^5$& 250 \\
\hline
0.2 & 3.01 & $6.04 \times 10^5$& 250 \\
\hline
0.3  & 3.61 & $7.25 \times 10^5$& 250 \\
\hline
0.4  & 4.89 & $9.83 \times 10^5$ & 250 \\
\hline
0.5 & 6.67 & $1.34 \times 10^6$ & 250 \\ 
\hline
0.6 & 8.71 & $1.76 \times 10^6$ & 250 \\
\hline
0.7  & 14.47 & $2.72 \times 10^6$ & 235.89 \\
\hline
0.8  & 14.16 & $1.89 \times 10^6$ & 180.98 \\
\hline
0.9  & 11.1 & $1.05 \times 10^6$ & 142.72 \\
\hline
1.0  & 11.65 & $1.02 \times 10^6$ &  135.32 \\
\hline
\end{tabular}
\label{tab:friction_metrics}
\end{table}

\section{Conclusion}
\label{sec:conclusion}
In this paper, we introduced an improved version of our fully-implicit and penalty-based frictional contact method, Implicit Contact Model. 
To test the performance of our contact model, we formulated an end-to-end simulation framework for the novel and difficult contact scenario of flagella bundling in viscous fluid.
For this contact problem, we showed that IMC has comparable performance to the state-of-the-art while maintaining faster convergence.
Furthermore, we showcased visually convincing frictional results in an imaginary viscous environment where friction is non-negligible.

For future work, we wish to improve upon the stability and robustness of our friction model. Despite the implicit formulation, the number of iterations necessary to converge starts to increase as $\mu$ increases.
Another interesting avenue of research is the use of deep learning to learn physics-based dynamics for simulation. Neural networks, when properly trained, have been known to be able to generate nearly identical outputs as numerical simulations while achieving orders of magnitude reduction in computation.
Thus, utilizing the computational efficiency and differentiability of neural networks while maintaining physical realism is a promising direction. 

\section{Acknowledgements}
\label{sec:acknowledgements}
We are grateful for financial support from the National Science Foundation (NSF) under award number CMMI-2101751 and IIS-1925360. M.K.J. is grateful for support from NSF (CAREER-2047663, CMMI-2053971).







\clearpage

\bibliographystyle{elsarticle-num}
\bibliography{ms}

\clearpage


\section*{Supplementary material}

# A Fully Implicit Method for Robust Frictional Contact Handling in Elastic Rods
# — Supplemental Information —

Dezhong Tong[*,a], Andrew Choi[*,b], Jungseock Joo[c], M. Khalid Jawed[†,a]

[a]*Department of Mechanical and Aerospace Engineering,*
[b]*Department of Computer Science,*
[c]*Department of Communication,*
*University of California, Los Angeles,*
*Los Angeles, California 90095, United States*

## Appendix A. Fluid Model

In this section, we formulate the relationship between velocity and hydrodynamic forces for each node in a discretized rod. We utilize Regularized Stokeslet Segments (RSS) [1] method to capture the viscous drag forces exerted on a slender rod moving in a viscous fluid as shown in Fig. 2(c). The Stokeslet, the primary Green's function of Stokes flow, represents solutions for the Stokes equations as a superposition of fundamental solutions. The methods of regularized Stokeslets have been implemented in numerous studies such as self-propelled microorganisms [2] [3], hyperactivated sperm motility [4] [5], and flagella bundling [6] [7].

For a particular choice of regularization [1], the relationship of the velocity at an evaluation point $\hat{\mathbf{x}}$ and the corresponding regularized force $\mathbf{f}$ exerted on a point $\mathbf{x}$ is the regularized Stokeslet:

$$8\pi\eta\mathbf{u}(\hat{\mathbf{x}}) = \left(\frac{1}{R} + \frac{\epsilon^2}{R^3}\right)\mathbf{f}(\mathbf{x}) + \frac{(\mathbf{f}(\mathbf{x}) \cdot \mathbf{r})\mathbf{r}}{R^3}, \tag{A.1}$$

where $\eta$ is the fluid viscosity; $\mathbf{r} = \hat{\mathbf{x}} - \mathbf{x}$; $R^2 = \|\mathbf{r}\|^2 + \epsilon^2$, and $\epsilon$ is the regularization parameter.

For an edge $\mathbf{e}^i = \mathbf{x}_{i+1} - \mathbf{x}_i$ in a viscous fluid, a point along this edge can be defined by $\mathbf{x}_\alpha = \mathbf{x}_i + \alpha(\mathbf{x}_{i+1} - \mathbf{x}_i)$, where $\alpha \in [0, 1]$. We assume a linear density of the force $\mathbf{f}_\alpha$ applied on the edge: $\mathbf{f}_\alpha = \mathbf{f}_i + \alpha(\mathbf{f}_{i+1} - \mathbf{f}_i)$. With this, we obtain the relationship between the velocity at $\hat{\mathbf{x}}$ and the linear force density as

$$8\pi\eta\mathbf{u}(\hat{\mathbf{x}}) = \|\mathbf{e}\| \int_0^1 \left[\left(\frac{1}{R_\alpha} + \frac{\epsilon^2}{R_\alpha^3}\right)\mathbf{f}_\alpha + \frac{(\mathbf{f}_\alpha \cdot \mathbf{r}_\alpha)\mathbf{r}_\alpha}{R_\alpha^3}\right] d\alpha, \tag{A.2}$$

---

[*] Equal contribution.
[†] Corresponding author. email: `khalidjm@seas.ucla.edu`

where $\|\mathbf{e}\|$ is the length of the edge $\mathbf{e}$. Through integration, we can then rewrite Eq. A.2 as

$$8\pi\eta\mathbf{u}(\hat{\mathbf{x}}) = \mathbf{f}_0(T_{0,-1} + \epsilon^2 T_{0,-3}) + \mathbf{f}_1(T_{1,-1} + \epsilon^2 T_{1,-3}) + \sum_{n=0}^{3} c_n T_{n,-3}, \quad (A.3)$$

where the coefficients $\mathbf{c}_n$ are

$$\begin{aligned}
\mathbf{c}_0 &= (\mathbf{f}_0 \cdot \mathbf{r}_0)\mathbf{r}_0, \\
\mathbf{c}_1 &= (\mathbf{f}_0 \cdot \mathbf{e})\mathbf{r}_0 + (\mathbf{f}_0 \cdot \mathbf{r}_0)\mathbf{e} + (\mathbf{f}_1 \cdot \mathbf{e})\mathbf{r}_0, \\
\mathbf{c}_2 &= (\mathbf{f}_0 \cdot \mathbf{v})\mathbf{v} + (\mathbf{f}_1 \cdot \mathbf{r}_0)\mathbf{e} + (\mathbf{f}_1 \cdot \mathbf{e})\mathbf{r}_0, \\
\mathbf{c}_3 &= (\mathbf{f}_1 \cdot \mathbf{e})\mathbf{e},
\end{aligned} \quad (A.4)$$

and the sequences of $T_{k,l}$ are

$$\begin{aligned}
\mathbf{T}_{0,-1} &= \frac{1}{\|\mathbf{e}\|} \log[\|\mathbf{e}\|R_\alpha + (\mathbf{r}_\alpha \cdot \mathbf{e})]\Big|_0^1, \\
\mathbf{T}_{0,-3} &= -\frac{1}{R_\alpha[\|\mathbf{e}\|R_\alpha + (\mathbf{r}_\alpha \cdot \mathbf{e})]}\Big|_0^1, \\
\mathbf{T}_{1,-1} &= \frac{R_\alpha}{\|\mathbf{e}\|^2}\Big|_0^1 - \frac{\mathbf{r}_0 \cdot \mathbf{e}}{\|\mathbf{e}\|^2} T_{0,-1}, \\
\mathbf{T}_{1,-3} &= -\frac{1}{R_\alpha\|\mathbf{e}\|^2}\Big|_0^1 - \frac{\mathbf{r}_0 \cdot \mathbf{e}}{\|\mathbf{e}\|^2} T_{0,-3}, \\
\mathbf{T}_{2,-3} &= -\frac{\alpha}{R_\alpha\|\mathbf{e}\|^2}\Big|_0^1 + \frac{1}{\|\mathbf{e}\|^2} T_{0,-1} - \frac{\mathbf{r}_0 \cdot \mathbf{e}}{\|\mathbf{e}\|^2} T_{1,-3}, \\
\mathbf{T}_{3,-3} &= -\frac{\alpha^2}{R_\alpha\|\mathbf{e}\|^2}\Big|_0^1 + \frac{2}{\|\mathbf{e}\|^2} T_{1,-1} - \frac{\mathbf{r}_0 \cdot \mathbf{e}}{\|\mathbf{e}\|^2} T_{2,-3}.
\end{aligned} \quad (A.5)$$

Recall that in DER a flexible rod constitutes $N$ nodes and $N-1$ edges each with length $\|\mathbf{e}^i\|$. Therefore, we can formulate hydrodynamic forces for the whole rod through the summation of Eq. A.3 for each segment, resulting in

$$8\pi\eta\mathbf{u}(\hat{\mathbf{x}}) = \sum_{i=0}^{N-2} \left( \mathbb{A}_1^i \mathbf{f}_i + \mathbb{A}_2^i \mathbf{f}_{i+1} \right), \quad (A.6)$$

where coefficient matrices $\mathbb{A}_1^i$ and $\mathbb{A}_2^i$ are

$$\begin{aligned}
\mathbb{A}_1^i &= \|\mathbf{e}^i\| \left[ \left(T_{0,-1}^{i,i+1} + \epsilon^2 T_{0,-3}^{i,i+1}\right) + T_{0,-3}^{i,i+1}\left(\mathbf{r}_i \mathbf{r}_i^T\right) + T_{1,-3}^{i,i+1}\left(\mathbf{r}_i(\mathbf{e}^i)^T + \mathbf{e}^i \mathbf{r}_i^T\right) + T_{2,-3}^{i,i+1}\left(\mathbf{e}^i(\mathbf{e}^i)^T\right) \right] - \mathbb{A}_2^i, \\
\mathbb{A}_2^i &= \|\mathbf{e}^i\| \left[ \left(T_{1,-1}^{i,i+1} + \epsilon^2 T_{1,-3}^{i,i+1}\right) + T_{1,-3}^{i,i+1}\left(\mathbf{r}_i \mathbf{r}_i^T\right) + T_{2,-3}^{i,i+1}\left(\mathbf{r}_i(\mathbf{e}^i)^T + \mathbf{e}^i \mathbf{r}_i^T\right) + T_{3,-3}^{i,i+1}\left(\mathbf{e}^i(\mathbf{e}^i)^T\right) \right].
\end{aligned} \quad (A.7)$$

Finally, by using Eqs. A.6 and A.7, we can describe the balance between the velocity and the force density along a discretized rod as the linear system

$$\begin{aligned}
\mathbf{U} &= \mathbb{A}\mathbf{f}, \\
\mathbf{F}^{\text{hydro}} &= [\mathbf{f}_0 \Delta l_0, \mathbf{f}_0 \Delta l_1, ..., \mathbf{f}_{N-1} \Delta l_{N-1}]^T,
\end{aligned} \quad (A.8)$$

2

where $\mathbf{U} = [\dot{\mathbf{x}}_0, \dot{\mathbf{x}}_1, ..., \dot{\mathbf{x}}_{N-1}]^T$ (the velocity of the liquid at nodal position $\mathbf{x}_i$ has the same speed of the rod due to non-slip boundary); $\mathbf{f} = [\mathbf{f}_0, \mathbf{f}_1, ..., \mathbf{f}_{N-1}]^T$ is the force density; $\mathbf{F}^{\text{hydro}}$ is the hydrodynamic force vector, and $\Delta l_i$ is the Voronoi length at node $\mathbf{x}_i$. We assume the density of the fluid to be equal to the density of the rod so that buoyant forces are negligible. As stated in Ref [1], we choose the regularization parameter $\epsilon = 1.02h$ based on the value of $\Delta l_i/h$ where $\Delta l_i$ is the Voronoi length.

**Appendix B. IMC Algorithmic Components**

*Appendix B.1. Scaling*

Since the radius of the rod may be very thin (in the range of a few mm), collision detection and gradient and Hessian computations may be susceptible to floating point inaccuracies. To mitigate this, we scale all nodes by a factor of $1/h$ for numeric stability. We denote $\bar{\Delta} = \Delta/h$ and $\bar{\delta} = \delta/h$ and as a result, normalize Eq. 9 to become

$$E\left(\bar{\Delta}, \bar{\delta}\right) = \begin{cases} (2 - \bar{\Delta})^2 & \bar{\Delta} \in (0, 2 - \bar{\delta}] \\ \left(\frac{1}{K_1} \log(1 + \exp(K_1(2 - \bar{\Delta})))\right)^2, & \bar{\Delta} \in (2 - \bar{\delta}, 2 + \bar{\delta}) \\ 0 & \bar{\Delta} \geq 2 + \bar{\delta}. \end{cases} \quad \text{(B.1)}$$

Accounting for this scaling in the gradient and Hessian computation can easily be taken care of through the product and chain rule which ends in us multiplying the gradient by $1/h$ and the Hessian by $(1/h)^2$. With this normalization in mind, we choose a stiffness $K_1$ for a desired $\delta$ through the relation

$$K_1(\delta, h) = \frac{15h}{\delta}. \quad \text{(B.2)}$$

*Appendix B.2. Collision Detection*

For collision detection, we simply obtain the set of all edge combinations whose minimum distance is less than $2h + \delta$, resulting in the contact set

$$C = \{\mathbf{x}_{ij} \in \mathcal{X} \mid \Delta_{ij} < 2h + \delta\}. \quad \text{(B.3)}$$

As this operation can be quite computationally expensive, we instead compute a candidate set $\hat{C} = \{\mathbf{x}_{ij} \in \mathcal{X} \mid \Delta_{ij} < 2h + \hat{\delta}\}$ at the beginning of each time step where $\hat{\delta} > \delta$ by a large enough margin. We then compute the actual contact set $C$ from the candidate set $\hat{C}$ at the start of each iteration, significantly reducing computational cost.

Note that if $\hat{\delta}$ is not set large enough, certain edge combinations not belonging to the initial set $\hat{C}$ may enter the contact zone or even penetrate by the end of the optimization. Our energy formulation in Eq. 9 is capable of dealing with this as minor penetrations do not lead to simulation failure and will be remedied in the next time step. This is in contrast to IPC, which may require a significantly larger $\hat{\delta}$ and/or more robust collision checking during each iteration of the optimization process.

*Appendix B.3. Contact Stiffness*

As a penalty method, a contact stiffness $k$ is used to scale the contact force and Jacobian. An appropriate value must be used to ensure that penetration (caused by too low of a value) or excessive hovering (caused by too large of a value) does not occur. First, let us denote the set of all $i$ and $j$ node indices of the contact set edge combinations $\mathbf{x}_{ij} \in C$ as the set $C_i$. Then, to generate an appropriate scaling for the contact stiffness, we compute the norm of the sum of forces (minus contact and friction) $\left\|\mathbf{F}_i^{\text{total}}\right\|$ experienced by each node in the contact set $C$. We can denote these forces as the set

$$\mathcal{F} = \left\{ \left\|\mathbf{F}_i^{\text{total}}\right\| \mid i \in C_i \right\}. \quad \text{(B.4)}$$

**Algorithm 1:** LineSearch

**Parameters:** $\alpha_l, \alpha_u, m_1 = 0.1, m_2 = 0.9$      // Initial interval for $\alpha$
**Input:** $\mathbf{q}, \Delta\mathbf{q}$      // DOFs from DER
**Output:** $\alpha$      // Newton search magnitude

1 **Function** LineSearch($\mathbf{q}, \Delta\mathbf{q}$):
2    iter $\leftarrow 0$
3    $\alpha \leftarrow 1$
4    success $\leftarrow$ False
5    $\mathbf{d_0} \leftarrow \mathbf{F}^T(\partial\mathbf{F}/\partial\mathbf{q})\Delta\mathbf{q}$
6    **while** success is False **do**
7      **if** $\alpha m_2 \mathbf{d_0} <= \frac{1}{2}\|\mathbf{F}(\mathbf{q} - \alpha\Delta\mathbf{q})\|^2 - \frac{1}{2}\|\mathbf{F}(\mathbf{q})\|^2 <= \alpha m_1 \mathbf{d_0}$ **then**
8        success $\leftarrow$ True
9      **else if** $\frac{1}{2}\|\mathbf{F}(\mathbf{q} - \alpha\Delta\mathbf{q})\|^2 - \frac{1}{2}\|\mathbf{F}(\mathbf{q})\|^2 < \alpha m_2 \mathbf{d_0}$ **then**
10       $\alpha_l \leftarrow \alpha$
11      **else**
12        $\alpha_u \leftarrow \alpha$
13      **if** $|\alpha_l - \alpha_r| <$ small value or iter $>$ iterMax **then**
14        success $\leftarrow$ True
15      $\alpha \leftarrow 0.5(\alpha_l + \alpha_u)$
16      iter $\leftarrow$ iter + 1
17    **end**
18    **return** $\alpha$

The maximum force magnitude of these forces can then be used to determine the contact stiffness

$$k = \max(\mathcal{F})s, \tag{B.5}$$

where $s$ is a constant scaling factor. In all our experiments, we set $s$ to be $1 \times 10^5$. The intuition behind this contact stiffness formulation is to achieve non-penetration through force equilibrium. Furthermore, by using the maximum of $\mathcal{F}$, if non-penetration can be achieved for the edge with the largest value in $\mathcal{F}$, then this $k$ value should be large enough to prevent penetration for all other contact pairs as well.

*Appendix B.4. Line Search*

Once the internal forces (e.g., bending force, twisting force, and stretching force), external forces (e.g., viscous dragging force and contact forces), and their respective Jacobians are computed, we can simply use Newton method to find the solution of the equations of motion. However, due to the high nonlinearity of the governing equations, convergence for Newton's method may suffer without a line search method. To rectify this, we perform Goldstein-Price line search in the Newton direction to ensure that the square of the Euclidean norm of total force $\|\mathbf{F}\|^2$ in Eq. 7 decreases.

We design an inner loop for each Newton iteration where we deploy the line search algorithm. This inner loop returns an optimal search step size $\alpha$ until $\|\mathbf{F}\|^2$ is smaller than a certain tolerance or until a maximum number of iterations is reached. As mentioned in Sec. 3.1, our energy formulation allows us to use a more aggressive line search strategy compared to Ref [8], resulting

in larger search step sizes and faster convergence. Pseudocode for the line search method can be found in detail in Appendix Algorithm 1.

*Appendix B.5. Full Pseudocode*

---

**Algorithm 2:** Flagella Simulation

**Parameters:** $\delta, \hat{\delta}, \nu$, tolerance
**Input:** $\mathbf{q}(t_i), \dot{\mathbf{q}}(t_i)$
**Output:** $\mathbf{q}(t_{i+1}), \dot{\mathbf{q}}(t_{i+1})$
**Require** : boundary conditions ← free

1 **Function** FlagellaSim($\mathbf{q}(t_i), \dot{\mathbf{q}}(t_i)$):
2     Guess $\mathbf{q}^{(0)} \leftarrow \mathbf{q}(t_i)$
3     $n \leftarrow 0, \epsilon \leftarrow \infty$
4     $\mathbf{F}^{\text{hydro}} \leftarrow \mathbb{A}^{-1} \dot{\mathbf{q}}(t_i) \Delta l$     // Eq. A.8
5     $\mathbf{F}^{\text{total}} \leftarrow \mathbf{0}^{\text{DOF}}$
6     $\mathbf{J}^{\text{total}} \leftarrow \mathbf{0}^{\text{DOF} \times \text{DOF}}$
7     **while** $\epsilon >$ tolerance **do**
8         $\mathbf{F}^{\text{int}} \leftarrow$ genForces($\cdot$)     // Eq. 6
9         $\mathbf{J}^{\text{int}} \leftarrow$ genJacobian($\cdot$)     // $\frac{\partial^2 (E_s + E_t + E_b)}{\partial q_i \partial q_j}$
10        **if** $n == 0$ **then**     // run only on first iter
11             $\hat{C} \leftarrow$ constructCandidateSet($\mathbf{x}, \hat{\delta}$)     // Appendix B.2
12             $k \leftarrow$ updateConStiffness($\hat{C}, \mathbf{F}^{\text{int}}$)     // Appendix B.3
13        $C \leftarrow$ collisionDetection($\hat{C}, \delta$)
14        **for** $\mathbf{x}, \mathbf{x}_0 \in C$ **do**
15             $\mathbf{F}^c, \mathbf{J}^c, \mathbf{F}^{\text{fr}}, \mathbf{J}^{\text{fr}} \leftarrow$ IMC($\mathbf{x}^{(n)}, \mathbf{x}_0, k, \delta, \nu$)     // Alg. 1
16             $\mathbf{F}^{\text{total}} \leftarrow \mathbf{F}^{\text{total}} + \mathbf{F}^c + \mathbf{F}^{\text{fr}}$
17             $\mathbf{J}^{\text{total}} \leftarrow \mathbf{J}^{\text{total}} + \mathbf{J}^c + \mathbf{J}^{\text{fr}}$
18        **end**
19        $\mathbf{F}^{\text{total}} \leftarrow \mathbf{F}^{\text{total}} + \mathbf{F}^{\text{int}} + \mathbf{F}^{\text{hydro}}$
20        $\mathbf{J}^{\text{total}} \leftarrow \mathbf{J}^{\text{total}} + \mathbf{J}^{\text{int}}$
21        $\mathbf{F}^{\text{free}} \leftarrow \mathbf{F}^{\text{der}}$(free)     // Downsize to only include free DOFs
22        $\mathbf{J}^{\text{free}} \leftarrow \mathbf{J}^{\text{der}}$(free, free)
23        $\Delta \mathbf{q}^{\text{free}} \leftarrow \mathbf{F}^{\text{free}} / \mathbf{J}^{\text{free}}$     // Solve $\mathbf{J}^{\text{free}} \Delta \mathbf{q}^{\text{free}} = \mathbf{F}^{\text{free}}$
24        $\alpha \leftarrow$ LineSearch($\mathbf{q}^{\text{free}}, \Delta \mathbf{q}^{\text{free}}$)     // Appendix Alg. 1
25        $\mathbf{q}^{(n+1)}$(free) $\leftarrow \mathbf{q}^{(n)}$(free) $- \alpha \Delta \mathbf{q}_{\text{free}}$
26        $\epsilon \leftarrow \left\| \mathbf{F}^{\text{free}} \right\|$     // update error
27        $n \leftarrow n + 1$
28     **end**
29     $\mathbf{q}(t_{i+1}) \leftarrow \mathbf{x}^{(n)}$
30     $\dot{\mathbf{q}}(t_{i+1}) \leftarrow (\mathbf{q}(t_{i+1}) - \mathbf{q}(t_i))/dt$
31     **return** $\mathbf{q}(t_{i+1}), \dot{\mathbf{q}}(t_{i+1})$

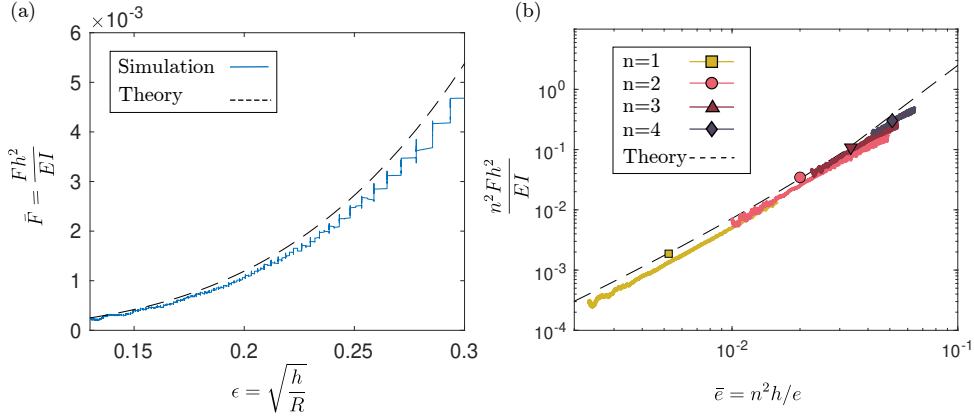

Figure C.1: Comparison of non-dimensionalized traction force $\bar{F}$ between IMC simulation results and theory for (a) tightening trefoil knots [9] and (b) tightening overhand knots of various unknotting numbers [11]. Simulation results are colored while theory is plotted as a dash line (C.1 for plot (a) and C.2 for plot (b)).

## Appendix C. Frictional Contact Force Validation

In this section, we provide validation for our generated frictional contact forces against theory concerning the contact scenario of knot tying. We first showcase validation against the theoretical relation for tightening trefoil knots formulated in [9] similar to our prior work in [10]. In addition to this, we also validate our contact model against the thoretical relation for tightening knots of varying unknotting numbers formulated in [11].

*Appendix C.1. Validation Study: Tightening a Trefoil Knot*

In Ref. [9], Audoly et al. formulated the theoretical relationship

$$\frac{Fh^2}{EI} = \frac{\epsilon^4}{2} + \mu\sigma\epsilon^3, \tag{C.1}$$

where $F$ is the traction force induced by friction; $h$ is the rod radius; $EI$ is the bending stiffness; $\mu$ is the coefficient of friction; $\sigma$ is a constant (0.492 for trefoil knots), and $\epsilon = \sqrt{h/R}$ where $R$ is the knot loop radius.

We use IMC to run a simulation of tightening a trefoil knot and record the traction forces $F$ for the parameters $h = 0.0016$ m, $E = 1.8e5$ Pa ($EI = E\pi h^4/4$), and $\mu = 0.1$. We non-dimensionalize traction force as $\bar{F} = Fh^2/EI$ and plot this against $\epsilon$. As shown in Fig. C.1(a), we observe excellent agreement between simulation results and the plotted theory.

*Appendix C.2. Validation Study: Tightening Knots of Various Unknotting Numbers*

In addition to the validation of [9], we also provide a more rigorous validation against theory for overhand knots of various unknotting numbers [11]. In this paper, Jawed et al. proposed the relation

$$n^2\frac{h}{e} = \frac{1}{8\sqrt{3}\pi^2} g\left(\left[\frac{96\sqrt{3}\pi^2}{\mu} \cdot \frac{n^2 Fh^2}{EI}\right]^{\frac{1}{3}}\right), \tag{C.2}$$

7

where $n$ is the unknotting number; $e$ is the end-to-end shortening, and $g(\cdot)$ is a known nonlinear equation detailed in the paper.

We use IMC to run simulations of tightening overhand knots for $n \in [1, 4]$ and $\mu = 0.07$ (all other rod parameters are the same as in Appendix C.1. We plot the recorded non-dimensionalized traction force $n^2 \bar{F}$ against the non-dimensionalized end-to-end shortening $\bar{e} = n^2 h/e$. Once again, we observe excellent agreement between simulation results and the plotted theory as shown in Fig. C.1(b), thus further cementing the physical accuracy of IMC's frictional contact forces.

## Appendix D. Parametric Study for Flagella Bundling

In this section, we perform a parametric study of flagella bundling using our simulation framework. Motivated by scaling analysis performed in Ref. [12], we choose four main dimensionless groups to examine: angular velocity $\bar{\omega} = \omega \nu z_0^4/EI$, fixed end distance $\Delta L/a$ (distance between neighboring flagella), rod number $M$, and friction coefficient $\mu$. We refer to all quantities with a ¯ as non-dimensionalized. Note that $\omega$ is the rotation velocity of the fixed end; $\nu$ is the viscosity of the fluid media; $z_0$ is the axial length of the flagella; $EI$ is the bending stiffness; $\Delta L$ is the distance between the neighboring fixed ends, and $a$ is the radius of helical geometry for one flagellum.

To evaluate the effects of our dimensionless groups, we define average propulsive force $\bar{F}_p$. When flagella bundle, propulsive forces are generated on each fixed end. We compute the average propulsive force on each fixed end during a fixed time period (defined as number of revolutions at constant $\omega$) and average this among the $M$ flagella to obtain $F_p$. With this, we can obtain

$$\bar{F}_p = \frac{F_p h^2}{EI}. \tag{D.1}$$

In Fig. D.2, we showcase the influence of each dimensionless group. All simulations used semicoiled helices (pitch / radius ∼ 5) as the flagella geometry. For all simulations not studying the effects of friction coefficient $\mu$, we consider frictionless contact ($\mu = 0$). In addition, simulations not studying number of flagella $M$ used $M = 2$ while those not studying fixed end distance used $\Delta L/a = 3$. Fig. D.2(a) studies the influence of rotation velocity with the aforementioned parameters. In Fig. D.2(b), the influence of fixed end distance is studied with normalized rotation velocity $\bar{\omega} = 3056$. Next, we look into the influence of flagella number in Fig. D.2(c1) and (c2) where we set normalized rotation velocity $\bar{\omega} = 611$ (to minimize inertial effects). Finally, in Fig. D.2(d1) and (d2), we study the influence of friction coefficient with rotation velocity $\bar{\omega} = 15280$ since we want to observe contact influence under extreme bundling.

Overall, we conclude the following preliminary results:

- Higher normalized rotation velocities result in higher propulsive forces.

- Larger normalized fixed end distances result in higher propulsive forces, but the influence of fixed end distance taper off as the distance grows large enough.

- More flagella result in larger total average propulsive forces but lower per flagella average propulsive forces.

- The propulsive force is nearly identical for $\mu \in [0, 0.3]$ (low friction cases) and then starts to exponentially increase as $\mu$ increases past 0.3. We observe a near monotonic increase in $\bar{F}_p$ as $\mu$ increases, barring one abnormality explained in Fig. D.2.

We provide additional insights into the frictional data in Fig.D.2(d1) and (d2). First, as $\mu$ grows, the sticking surface between flagella grows, eventually resulting in "one" flagellum. This essentially results in a single flagellum with a higher effective stiffness, which translates to higher thrust forces as a single flagellum has been shown to have higher propulsive efficiency compared to multiple flagella with equal cumulative volume [13].

Overall, we showcase interesting preliminary results for flagella bundling provided entirely by our simulation framework. We emphasize here that presented results have not been validated with experiments. However, the results are quite intuitive and in particular, $\bar{\omega}$ and $\Delta L/a$ have similar tendencies with prior works [12].

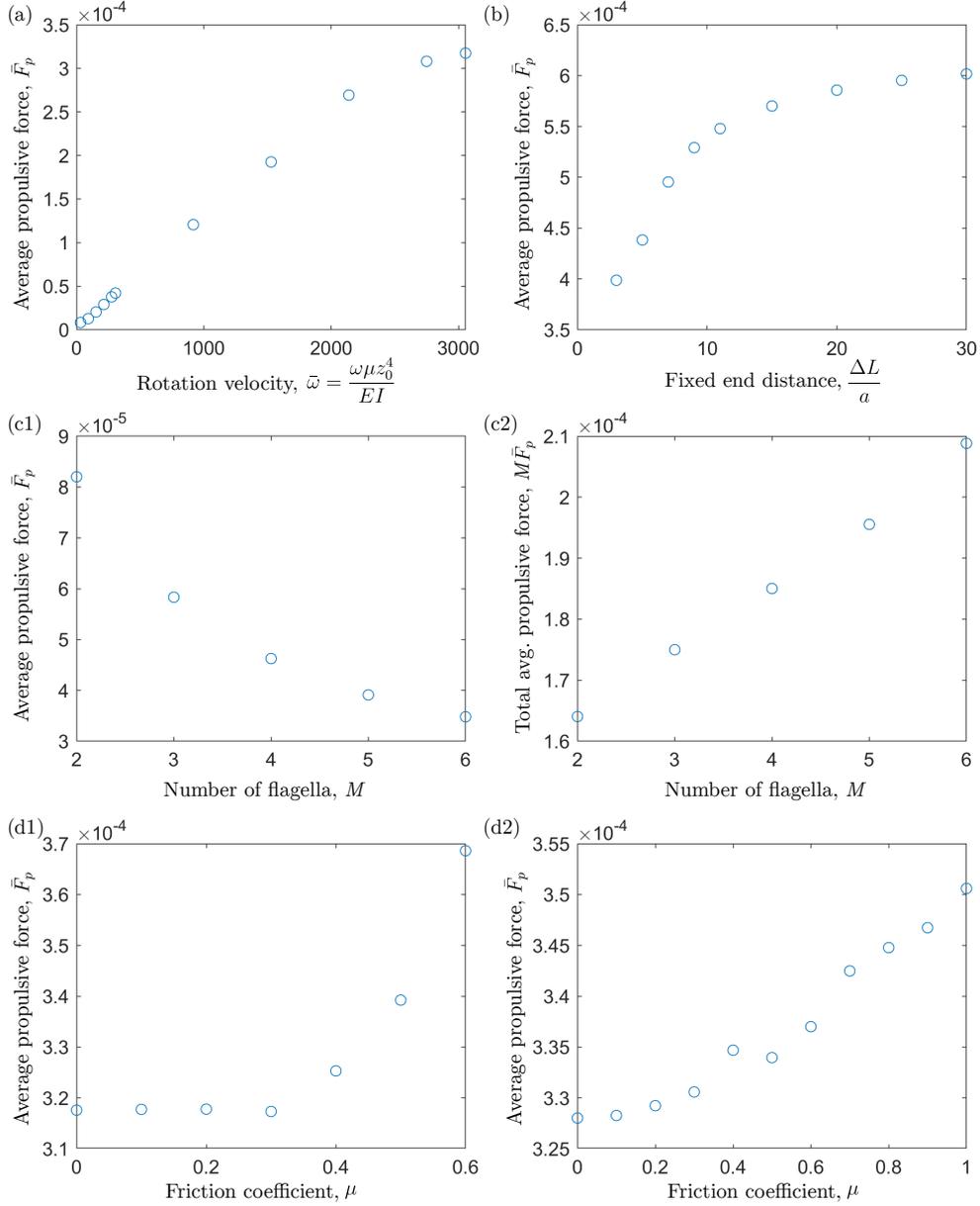

Figure D.2: Simulation results showcasing the relationship between average propulsive force $\bar{F}_p$ and (a) rotation velocity $\bar{\omega}$, (b) fixed end distance $\Delta L/a$, (c1) number of flagella $M$, and (d1) friction coefficient $\mu$ over 597 revolutions. In addition, we show (c2) the total average propulsive force $M\bar{F}_p$ against the number of flagella $M$ over 597 revolutions. Note that (d1) only includes $\mu \in [0, 0.6]$ as simulations with $\mu > 0.6$ ended before the total 597 revolutions. Instead, we plot (d2) the average propulsive force $\bar{F}_p$ against the full friction coefficient range $\mu \in [0, 1.0]$ over 322 revolutions (the earliest exit sim time). We note that the abnormal change between $\mu = 0.4$ and $\mu = 0.5$ for plot (d2) can be attributed to insufficient data as 322 revolutions is nowhere near the simulation end of 597 revolutions.

## Appendix E. Miscellaneous Information

*Appendix E.1. Stability of IMC and IPC*

We plot the cumulative iteration and computation time for both IMC and IPC for all flagella cases with respect to simulation time in Fig. E.3. When observing the plots, we can see that both methods have near linear growth rates (a sign of stability) for all metrics with IMC consistently possessing higher computational efficiency as indicated by the lower slopes.

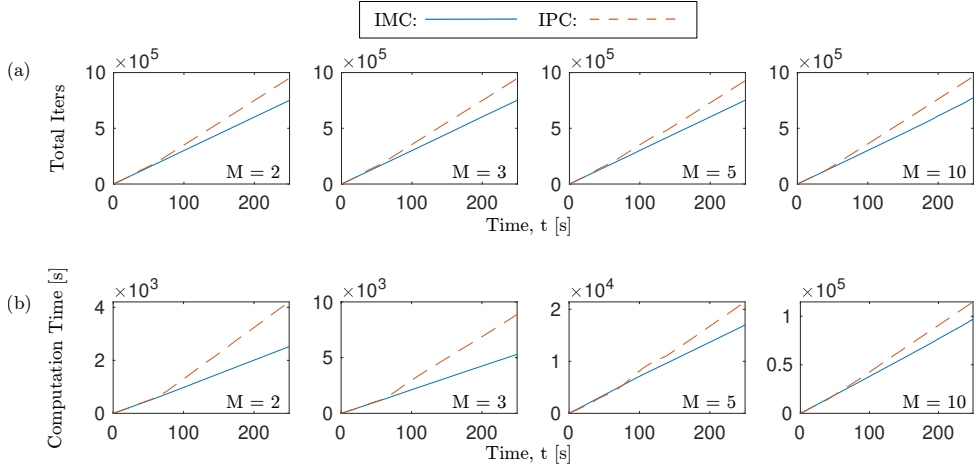

Figure E.3: Cumulative iteration and computation time with respect to simulation time.

*Appendix E.2. Solver and Computational Tools*

On the computational end, we use the sparse matrix solver oneMKL PARDISO from Intel's oneAPI Math Kernel Library [14] and optimized all operations by compiling Eigen to use MKL as the BLAS and LAPACK backend. All necessary gradients and Hessians were computed symbolically and created into callable functions optimized through LLVM using the open-source C++ library SymEngine [15]. All experiments were run single-threaded using an AMD Ryzen Threadripper 2990WX 32-core processor. Although parts of our code can be improved by multithreading, we decided to forgo this as solving the dense matrix in Eq. A.8 soon becomes a bottleneck.



\end{document}